\documentclass[aps, pra, twocolumn, superscriptaddress, tightenlines, longbibliography]{revtex4-1}

\usepackage{graphicx}
\usepackage{amsmath}
\usepackage{booktabs}
\usepackage{subfigure}
\usepackage{bm}
\usepackage{natbib}
\usepackage[colorlinks=true,linkcolor=blue,citecolor=blue,urlcolor=blue]{hyperref}

\begin{document}

	\title{Universal quantum gates by nonadiabatic holonomic  evolution for the surface electron}
	
	\author{Jun Wang}
	\affiliation{Department of Physics, Applied Optics Beijing Area Major Laboratory, Beijing Normal University, Beijing 100875, China}
	
	\author{Wan-Ting He}
	\affiliation{Department of Physics, Applied Optics Beijing Area Major Laboratory, Beijing Normal University, Beijing 100875, China}
	
	\author{Hai-Bo Wang}
	\email{hbwang@bnu.edu.cn}
	\affiliation{Department of Physics, Applied Optics Beijing Area Major Laboratory, Beijing Normal University, Beijing 100875, China}
	
	\author{Qing Ai}%
	\email{aiqing@bnu.edu.cn}
	\affiliation{Department of Physics, Applied Optics Beijing Area Major Laboratory, Beijing Normal University, Beijing 100875, China}

	\date{\today}

	\begin{abstract}
		The nonadiabatic holonomic quantum computation based on the geometric phase is robust against the built-in noise and decoherence. In this work, we theoretically propose a scheme to realize nonadiabatic holonomic quantum gates in a surface electron system, which is a promising two-dimensional platform for quantum computation. The holonomic gate is realized by a three-level structure that combines the Rydberg states and spin states via an inhomogeneous magnetic field. After a cyclic evolution, the computation bases pick up different geometric phases and thus perform a geometric gate. Only the electron with spin up experiences the geometric gate, while the electron with spin down is decoupled from the state-selective driving fields. The arbitrary controlled-$U$ gate encoded on the Rydberg states and spin states can then be realized. The fidelity of the output state exceeds 0.99 with experimentally achievable parameters.
	\end{abstract}
	
	\maketitle

	\section{Introduction}\label{sec1}
	
	The quantum geometric phase is a very important resource for quantum computation \cite{barry1983prl, berry1984prsl, leek2007science, ai2009pra, chitambar2019rmp, zhang2023pr}. Quantum gates based on the geometric phase are robust against the disturbance of the dynamic process owing to their global geometric properties. The adiabatic holonomic quantum computation (AHQC) realizes high-fidelity quantum gates via the geometric phase in an adiabatic evolution \cite{zanardi1999pla, osterloh2002nature, tong2004prl,yi2004prl, song2016njp,song2016pra}. The AHQC protocol is solely determined by the solid angle of the cyclic evolution in the parameter space, and thus is robust against small perturbations of the evolution path. The AHQC schemes have been proposed in various physical systems, such as superconducting qubits \cite{faoro2003prl}, trapped ions \cite{duan2001science}, and semiconductor quantum dots \cite{solinas2003prb}.
	
	However, the adiabatic condition of AHQC requires a long evolution time, which accumulates considerable decoherence. Thus, the nonadiabatic holonomic quantum computation (NHQC) was proposed \cite{wang2001prl, zhu2002prl, sjoqvist2012njp, xu2012prl, xu2014pra, xue2015pra, xu2015pra, zheng2016pra, xu2017pra, xue2017prap, chen2018adp, jin2023arxiv}. The NHQC preserves the computational universality of the AHQC but does not require the adiabatic condition, and thus has attracted broad interest in recent years \cite{zu2014nature, arroyo2014nc, yalenp2016np, sekiguchi2017np, abdumalikov2013nature, feng2013prl, zhou2017prl, leroux2018nc, nagata2018nc, xu2018prl, huang2019prl, yan2019prl, liu2019prl, ramberg2019prl, xu2020prl}.
	
	On the other hand, the electron on the surface of liquid helium provides a controllable two-dimensional (2D) quantum system, where the surface electron (SE) is attracted by the induced image charge inside the liquid helium and concurrently repulsed by the helium atoms. The confinement perpendicular to the surface leads to a hydrogen-like spectrum which can be used in quantum information tasks \cite{platzman1999science, wang2023adp}. Meanwhile, the motion parallel to the surface is free of defects and impurities, thus the SE forms a perfect 2D electron system which is widely observed in semiconductor devices \cite{kawakami2021prl}. The SE can be manipulated by the circuit QED architecture \cite{koolstra2019natcom, zhou2022nature} or the microchannel devices \cite{glasson2001prl, ikegami2009prl, rees2011prl, ikegami2012prl, rees2016prl, rees2016prb, badrutdinov2020prl} with high transport efficiency \cite{bradbury2011prl, kielpinski2022nature}. With a static magnetic field perpendicular to the surface, the motion parallel to the surface is quantized as orbital states \cite{chen2018prb}, which is similar to the Landau levels. In addition, the spin state of the SE is also an important quantum resource owing to its long relaxation time that exceeds 100 s \cite{lyon2006pra}. Both the Rydberg and orbital states can be coupled to the spin states of electrons \cite{kawakami2019prl, schuster2010prl, kawakami2023arxiv}. Recent works \cite{tokura2006prl, kawakami2023arxiv} show a practical method to couple the Rydberg state with the spin state using a local inhomogeneous magnetic field, where the electrons with different expected positions experience different magnetic fields depending on their Rydberg state.

	In the past, research on the SE has mainly focused on exploring the physical properties, while the efficient quantum gates based on the SE still require further investigation. The Rydberg states of SE can probably realize large-scale quantum computation owing to the long-range dipole-dipole interaction of adjacent electrons\cite{kawakami2023arxiv}, while the spin states possess long lifetime which may be valuable for quantum memory. Therefore, in this work, we propose a scheme to realize the nonadiabatic holonomic gates on both the spin states and Rydberg states of a single SE. We first propose an arbitrary single-qubit holonomic gate on the Rydberg state of the SE, which is realized by a three-level structure driven by time-dependent microwave pulses. During the cyclic evolution, two orthogonal bases pick up different geometric phases. The universal single-qubit holonomic gate is realized by varying the complex ratio between the Rabi frequencies of the two driving fields. Then we introduce an inhomogeneous magnetic field through a magnetized ferromagnetic electrode. The Rydberg states with different expected positions experience different magnetic fields, and thus their Zeeman energy splittings are different. By applying the state-selective pulses, three Rydberg states with spin up are coupled with the driving fields, while the Rydberg states with spin down are decoupled. An arbitrary holonomic single-qubit gate $U$ is applied on the three coupled states, while the three decoupled states remain unchanged. In this way, the holonomic controlled-$U$ gate of the Rydberg and spin states is achieved. Owing to the global geometric properties, the NHQC gates are not sensitive to the fluctuation of the pulse duration. Because the adiabatic condition is not required during the evolution, the fast manipulation makes the scheme robust against dissipation. Our theoretical scheme is based on the experimental configuration, and the parameters are experimentally achievable.
	
	This paper is organized as follows. In Sec.~\ref{sec2}, we introduce the basic model of the SE in the external magnetic and electric fields. In Sec.~\ref{sec4}, we present the holonomic CNOT gate scheme. Finally, we conclude the work and give a prospect in Sec.~\ref{sec5}. In Appendix~\ref{AppA}, we solve the wavefunctions in the electric holding field. In Appendix~\ref{AppB}, we investigate the relation between the decay rates and the electric field.

\section{The model}
\label{sec2}

\begin{figure}[!ht]
\centering
\includegraphics[width=1\linewidth]{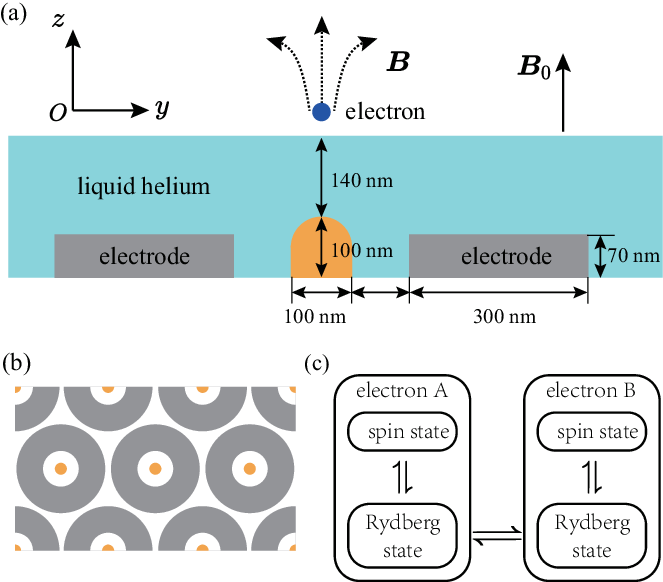}
\caption{(a) The longitudinal section of the proposed model. A central pillar electrode (orange) with a positive voltage of 60~mV and an annular electrode (dark gray) with a negative voltage of -45~mV are embedded in the liquid helium. The SE floats above the liquid surface. The vertical motion of the SE is quantized as the Rydberg states owing to the image potential and the electric field applied by the electrodes. A uniform static magnetic field $\boldsymbol{B}_0$ is perpendicularly applied to the surface. The center electrode is made of ferromagnetic material and induces an inhomogeneous magnetic field. The total magnetic field is $\boldsymbol{B}$. The motion in the $xOy$ plane is quantized as the Landau level induced by the vertical magnetic field and the cylindrically symmetric electric potential. (b) The top view of the architecture. The electrodes are arranged in an array, with one electron in each unit. Because the energy levels of each electron can be tuned independently by the voltage of electrodes, the driving field can be resonant with only one electron at a time. The adjacent electrons are coupled by the dipole-dipole interaction of the Rydberg states. (c) The quantum information can be stored in the spin states and manipulated in large-scaled quantum computation via the Rydberg states. }\label{fig1}
\end{figure}

Our proposal is based on the experimental configuration \cite{kawakami2023arxiv}. As shown in Fig.~\ref{fig1}, a pillar electrode with positive voltage and an annular electrode with negative voltage are embedded in the liquid helium. These electrodes apply an electric holding field with the cylindrically symmetric electric potential $V(r,z)$, where $z$ is the vertical coordinate and $r=\sqrt{x^2+y^2}$ is the radial coordinate. Meanwhile, the electron is confined by the image potential $-\Lambda e^2/z$ introduced by the image charge in the liquid helium,  where $e$ is the charge of the electron and $\Lambda=(\epsilon-1)/[4(\epsilon+1)]$ with the dielectric constant $\epsilon\approx 1.057$. The vertical motion of the SE is quantized as the Rydberg states. On the other hand, a uniform static magnetic field $\boldsymbol{B}_0=B_0\boldsymbol{e}_z$ is perpendicularly applied to the surface. Because the total potential $-\Lambda e^2/z-eV(r,z)$ is cylindrically symmetric, we introduce the symmetric gauge $\boldsymbol{A}=\boldsymbol{B}_0\times\boldsymbol{r}/2$. The motion of the electron is determined by the Hamiltonian
\begin{align}
H_T=&\frac{(\boldsymbol{p}+e\boldsymbol{A})^2}{2m_e}-\frac{\Lambda e^2}{z}-eV(r,z) \nonumber \\
=&H_0+\frac{1}{2}\omega_c L_z,
\end{align}
where $\omega_c=eB_0/m_e$ is the cyclotron frequency, $m_e$ is the mass of the electron, and $L_z$ is the angular momentum along $z$-direction, i.e.,
\begin{align}
	L_z=&xp_y-yp_x=-i\hbar(x\frac{\partial}{\partial y}-y\frac{\partial}{\partial x})=-i\hbar\frac{\partial}{\partial \phi},
\end{align}
where $p_\alpha$ and $\alpha$ are the momentum and position of the electron ($\alpha=x,y,z$), and $\phi$ is the azimuthal coordinate in the $xOy$-plane. The vertical and radial motion of the electron is determined by
\begin{align}
	H_0=&\frac{1}{2m_e}(p_x^2+p_y^2+p_z^2)+\frac{1}{8}m_e\omega_c^2(x^2+y^2) \nonumber \\
	&-\frac{\Lambda e^2}{z}-eV(r,z) \nonumber \\
	=&-\frac{\hbar^2}{2m_e}\Big(\frac{\partial^2}{\partial r^2}+\frac{1}{r}\frac{\partial}{\partial r}-\frac{m^2}{r^2}+\frac{\partial^2}{\partial z^2}\Big)+\frac{1}{8}m_e\omega_c^2r^2\nonumber \\
	&-\frac{\Lambda e^2}{z}-eV(r,z),
\end{align}
where $m$ is an integer. Because of the cylindrical symmetry, the wavefunction can be expressed as
	\begin{align}
		\Psi(z,r,\phi)=\psi_{n_z,n_r,m}(z,r)\Phi(\phi),
	\end{align}
where $\psi_{n_z,n_r,m}(z,r)$ is the wavefunction of $H_0$ with the vertical quantum number $n_z$, the radial quantum number $n_r$, and the angular quantum number $m$. $\Phi(\phi)=e^{im\phi}$ is the azimuthal wavefunction that satisfies $L_z \Phi(\phi)=m\hbar\Phi(\phi)$.

The vertical motion of the SE is quantized by the Rydberg state labeled by $n_z$. The expected positions of the lowest three states along the $z$ direction are 7.63 nm, 17.2 nm, and 25.3 nm, which are derived from the numerical solution of the wavefunction, cf. Appendix~\ref{AppA}. An inhomogeneous magnetic field is induced by the center electrode which is made of ferromagnetic material. The electrons with different $n_z$ have different expected positions $\langle z\rangle$, and thus experience different magnetic fields. The differences of the magnetic field $\Delta B$ at the expected positions with $n_z=1,2,3$ are on the order of 5~mT \cite{kawakami2023arxiv}. The corresponding difference of the Zeeman energy is $g\mu_B\Delta B/\hbar\sim0.88$~GHz, where $g$ is the Lande g factor and $\mu B=e\hbar/(2m_e)$ is the Bohr magneton. This energy difference is much larger than the decay rates of the Rydberg states, cf. Appendix~\ref{AppB}, and plays a significant role in the following controlled-$U$ gate scheme. 


\section{Nonadiabatic Holonomic Quantum Gates Based on The Rydberg states and spin states}
\label{sec4}

	\begin{figure}[!ht]
		\centering
		\includegraphics[width=1\linewidth]{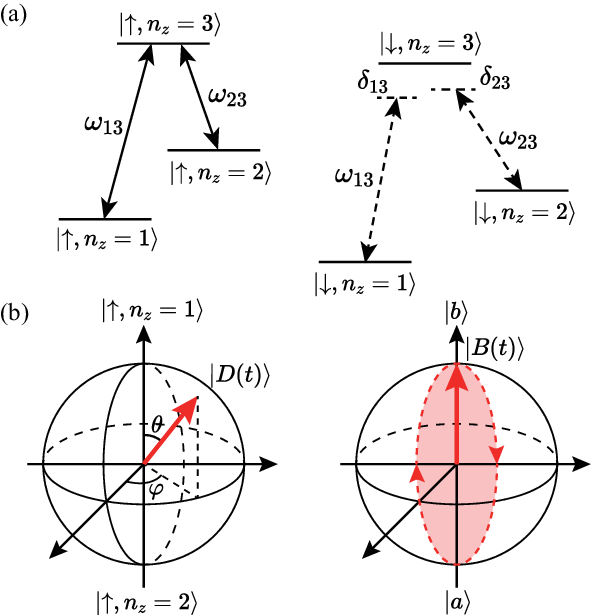}
		\caption{(a) Schematic diagram of the nonadiabatic holonomic gate. Two driving pulses are resonant with $|\uparrow,n_z=1\rangle \Leftrightarrow |\uparrow,n_z=3\rangle$ and $|\uparrow,n_z=2\rangle \Leftrightarrow |\uparrow,n_z=3\rangle$, respectively. The driving pulses are off-resonant with the transition frequencies of the spin-down states due to the large detuning $\delta_{13}$ and $\delta_{23}$. (b) The evolutions of the instantaneous orthogonal bases $|D(t)\rangle$ and $|B(t)\rangle$ on the Bloch spheres. The dark state $|D(t)\rangle$ remains unchanged, while $|B(t)\rangle$ evolves along the longitude circle and acquires a geometric phase. }\label{fig2}
	\end{figure}

At first, we will demonstrate the proposal to realize a single-qubit gate in the subspace of a specific spin state, that is, the subspace spanned by $\{|\uparrow,n_z\rangle \}$ ($n_z=1,2,3$). As we have mentioned in Sec.~\ref{sec2}, the Rydberg states with larger quantum number $n_z$ are further away from the liquid surface, and thus experience different magnetic fields. Since the Zeeman energy of the spin state is determined by the inhomogeneous magnetic field, the Zeeman energies of different Rydberg states are different. The magnetic field experienced by state $|n_z\rangle$ is $B_z^{(n_z)}$. The Zeeman-energy splitting between $|\uparrow,n_z\rangle$ and $|\downarrow,n_z\rangle$ is $g\mu_B B_z^{(n_z)}$, where $\uparrow$ and $\downarrow$ represent the spin-up and spin-down states, respectively. As shown in Fig.~\ref{fig2}(a), if we label the transition frequency of $|\uparrow,n_z\rangle\Leftrightarrow |\uparrow,3\rangle$ as $\omega_{n_z3}$ ($n_z=1,2$), then the transition frequency of $|\downarrow,n_z\rangle\Leftrightarrow |\downarrow,3\rangle$ is $\omega_{n_z3}+\delta_{n_z3}$, with $\delta_{n_z3}=g\mu_B[B_z^{(n_z)}-B_z^{(3)}]$.	Therefore, by applying two state-selective driving fields with frequency $\omega_{n_z3}$, the three Rydberg states with spin up form a $\Lambda$-type three-level structure, while the Rydberg states with spin down are decoupled. The driving pulses are controlled by an arbitrary-waveform generator. The Rabi frequencies of $|\uparrow,1\rangle \Leftrightarrow |\uparrow,3\rangle$ and $|\uparrow,2\rangle \Leftrightarrow |\uparrow,3\rangle$ are respectively $\Omega_1(t)$ and $\Omega_2(t)$. The ratio between $\Omega_1(t)$ and $\Omega_2(t)$ is a constant, i.e., $\Omega_1(t)=\Omega(t)\sin(\theta/2)e^{i\varphi}$ and $\Omega_2(t)=-\Omega(t)\cos(\theta/2)$ with $\Omega(t)=\sqrt{\Omega_1^2(t)+\Omega_2^2(t)}$. The interaction Hamiltonian reads
\begin{align}
	H_I(t)=&~\Omega(t)\Big(\sin\frac{\theta}{2}e^{i\varphi}|\uparrow,3\rangle\langle \uparrow,1| \nonumber\\ &-\cos\frac{\theta}{2}|\uparrow,3\rangle\langle \uparrow,2|+\textrm{H.c.}\Big). \label{eqHI}
\end{align}
Hereafter, we assume $\hbar=1$ for simplicity.  $\Omega(t)$ represents the shape of the driving pulse. The duration of the driving pulse can be very short because the adiabatic condition is not required during the evolution. A specific pulse shape is not strictly required for NHQC, but the integral over time needs to be $\pi$, i.e., $\int_{-\infty}^{\infty}\Omega(t)dt=\pi$, which will be explained later.
	
The eigenenergies of $H_I$ are $0,~\pm\Omega$. The corresponding eigenstates are
	\begin{align}
		|\psi_0\rangle=&~|d\rangle, \\
		|\psi_+\rangle=&~\frac{1}{\sqrt{2}}(|b\rangle+|a\rangle), \\
		|\psi_-\rangle=&~\frac{1}{\sqrt{2}}(|b\rangle-|a\rangle),
	\end{align}
where $|d\rangle\equiv\cos(\theta/2)|\uparrow,1\rangle+\sin(\theta/2)e^{i\varphi}|\uparrow,2\rangle$ is the dark state,  $|b\rangle\equiv \sin(\theta/2)e^{-i\varphi}|\uparrow,1\rangle-\cos(\theta/2)|\uparrow,2\rangle$ is the bright state, and $|a\rangle=|\uparrow,3\rangle$ is the intermediate state. The dark state does not evolve with time because the corresponding eigenenergy is zero. Thus, we define
\begin{align}
	|D(t)\rangle\equiv U_1(t)|d\rangle=|d\rangle,
\end{align}
where
\begin{align}
	U_1(t)=&~\mathcal{T}\exp[i\int_0^{t}H_I(t')dt'],
\end{align}
is the evolution operator of $H_I$, and $\mathcal{T}$ represents the time-ordered integration. We also define
\begin{align}
		|B(t)\rangle\equiv&~e^{i\alpha(t)}U_1(t)|b\rangle\nonumber \\
		=&~e^{i\alpha(t)}[\cos\alpha(t)|b\rangle-i\sin\alpha(t)|a\rangle],
	\end{align}
where $\alpha(t)=\int_0^{t}\Omega(t')dt'$. Here we introduce a global phase $e^{i\alpha(t)}$ to ensure a cyclic evolution of $|B(t)\rangle$ when $\alpha=\pi$ at the final time. The evolutions of $|D(t)\rangle$ and $|B(t)\rangle$ are shown in Fig.~\ref{fig2}(b). The state $|D(t)\rangle$ is unchanged, while the state $|B(t)\rangle$ evolves along the longitude line of the Bloch sphere with bases $\{|b\rangle, |a\rangle\}$ and induces a geometric phase. To make use of the geometric phase, we introduce the following instantaneous orthogonal bases,
	\begin{align}
		&|\xi_1(t)\rangle\equiv \sin\frac{\theta}{2}e^{i\varphi}|B(t)\rangle+\cos\frac{\theta}{2}|D(t)\rangle, \\
		&|\xi_2(t)\rangle\equiv -\cos\frac{\theta}{2}|B(t)\rangle+\sin\frac{\theta}{2}e^{-i\varphi}|D(t)\rangle.
	\end{align}
By choosing $\alpha(\tau)=\pi$ at the final time $\tau$,  $|\xi_1(\tau)\rangle=|\xi_1(0)\rangle=|\uparrow,1\rangle$ and $|\xi_2(\tau)\rangle=|\xi_2(0)\rangle=|\uparrow,2\rangle$, i.e., $|\xi_1(t)\rangle$ and $|\xi_2(t)\rangle$ coincide with the computation bases $|\uparrow,1\rangle$ and $|\uparrow,2\rangle$ both at the beginning and end of time. It can be easily verified that the parallel transport condition $\langle\xi_1(t)|H_I(t)|\xi_2(t)\rangle=0$ is satisfied during the whole evolution $t\in[0,\tau]$. Thus, the dynamic phases vanish and the evolution operator $U(\tau)$ in the subspace spanned by $|\uparrow,1\rangle$ and $|\uparrow,2\rangle$ is \cite{sjoqvist2012njp, zu2014nature}
\begin{align}
		U(\tau)=&~\mathcal{T}\exp\Big[i\int_0^{\tau}\mathcal{A}(t)dt\Big] \nonumber \\
		=&~\begin{pmatrix}
			\cos\theta & e^{-i\varphi}\sin\theta \\
			e^{i\varphi}\sin\theta & -\cos\theta
		\end{pmatrix} \nonumber\\
	=&~\boldsymbol{n}\cdot\boldsymbol{\sigma},\label{eqU}
\end{align}
where $\boldsymbol{n}=(\sin\theta\cos\varphi, \sin\theta\sin\varphi, \cos\theta)$, $\boldsymbol{\sigma}=(\sigma_x, \sigma_y, \sigma_z)$ are the Pauli operators, and the connection matrix $\mathcal{A}$ is
\begin{align}
	\mathcal{A}=-\dot{\alpha}\begin{pmatrix}
		\sin^2\frac{\theta}{2} & -\sin\frac{\theta}{2}\cos\frac{\theta}{2}e^{-i\varphi} \\[3pt]
		-\sin\frac{\theta}{2}\cos\frac{\theta}{2}e^{i\varphi} & \cos^2\frac{\theta}{2}
	\end{pmatrix}
\end{align}
whose matrix elements are determined by
\begin{align}
	\mathcal{A}_{ij}=\langle\xi_i(t)|i\partial_t|\xi_j(t)\rangle.
\end{align}
Therefore, a single-qubit holonomic gate on the coupled Rydberg states is realized by adjusting the complex ratio $\tan(\theta/2)e^{i\varphi}$ between the Rabi frequencies of the two driving fields. For example, a Hadamard gate $H$ is realized by $(\theta,\varphi)=(\pi/4,0)$, and a NOT gate $X$ is realized by $(\theta,\varphi)=(\pi/2,0)$. In addition, two sequential holonomic gates lead to
\begin{align}
	U^{(m)}U^{(n)}=\boldsymbol{n}\cdot\boldsymbol{m}-i\boldsymbol{\sigma}\cdot(\boldsymbol{n}\times\boldsymbol{m}), \label{eqU2}
\end{align}
which forms an arbitrary SU(2) transformation that rotates the state around the axis $\boldsymbol{n}\times\boldsymbol{m}$ by the angle $2\arccos(\boldsymbol{n}\cdot\boldsymbol{m})$
\cite{nielsen2010, sjoqvist2012njp}. For instance, the $\pi/8$ phase gate \cite{nielsen2010} is realized by two sequential gates with $(\theta,\varphi)=(\pi/2,0)$ and ($\theta,\varphi)=(\pi/2,\pi/8)$.

Next, we will demonstrate the two-qubit gate proposal by taking two spin states into account. As shown in Fig.~\ref{fig2}(a), the Rydberg states with spin down are off-resonant with the driving fields. Thus, the subspace spanned by $\{|\downarrow,n_z\rangle \}$ ($n_z=1,2,3$) is decoupled with the driving fields. The total evolution operator in the subspace spanned by $\{|\downarrow,1\rangle, |\downarrow,2\rangle, |\uparrow,1\rangle, |\uparrow,2\rangle\}$ is
\begin{align}
	U_{\textrm{tot}}(\tau)=|\downarrow\rangle\langle\downarrow|\otimes I +|\uparrow\rangle\langle\uparrow|\otimes U,
\end{align}
where $U$ is the single-qubit gate on the Rydberg states according to Eq.~(\ref{eqU}). In this way, the holonomic controlled-$U$ gate with the spin state being the control qubit is realized. In addition, according to Eq.~(\ref{eqU2}), by applying two controlled gate sequentially, we can realize an arbitrary controlled-$U$ gate. For example, by choosing $(\theta,\varphi)=(\pi/2,0)$, $U=X$ and a CNOT gate is realized as
\begin{align}
	U_{\textrm{tot}}(\tau)=|\downarrow\rangle\langle\downarrow|\otimes I +|\uparrow\rangle\langle\uparrow|\otimes X,
\end{align}
where $I$ and $X$ are the identity operator and qubit-flip operator in the subspace spanned by $\{|n_z=1\rangle, |n_z=2\rangle\}$, respectively. The Rydberg states flip only if the spin state is $|\uparrow\rangle$. 

In the presence of dissipation, the evolution of the system can be described by the quantum master equation \cite{Breuer2002}
	\begin{align}
		\frac{\partial}{\partial t}\rho=-i[H,\rho]-\mathcal{L}(\rho),
	\end{align}
where the Lindblad operator is
	\begin{align}
		\mathcal{L}(\rho)=~&\sum_{m,n}\kappa_{mn}[C_{mn} \rho C^\dagger_{mn} -\frac{1}{2}\{C^\dagger_{mn}C_{mn},\rho\}], \label{eq7}
	\end{align}
where $\{A,B\}=AB+BA$ is the anti-commutator, $C_{mn}=|m\rangle\langle n|$ is the collapse operator with the corresponding decay rate $\kappa_{mn}$. It is noteworthy that $\kappa_{mn}$ increases with the electric holding field $E_{\perp}$ induced by the electrodes, cf. Appendix~\ref{AppB}. For typical experimental configuration, $E_{\perp}$ is on the order of $100\sim1000$~V/cm \cite{zou2022njp}. Hereafter the evolution with dissipation is solved by QuTiP \cite{Johansson2012CPC, Johansson2013CPC}.
		
The state fidelity $F$ between the final state $\rho(t)$ and the ideal target state $\rho_i$ is defined as \cite{nielsen2010}
	\begin{align}
		F=\textrm{Tr}\sqrt{\sqrt{\rho_i}\rho(t)\sqrt{\rho_i}}. \label{eqF}
	\end{align}
	
	\begin{table}[!h]
		\centering
		\renewcommand\arraystretch{1.5}
		\tabcolsep=0.3cm
		\caption{The output-state fidelity of the CNOT gate under typical input states. The decay rates are calculated under a typical electric holding field $E_{\perp}=100$~V/cm. }
		\begin{tabular}{c c l}
			\toprule[1.5pt]
			Input state & Ideal output state & Fidelity \\
			\midrule[1pt]
			$|\downarrow,1\rangle$ & $|\downarrow,1\rangle$ & 1 \\
			$|\downarrow,2\rangle$ & $|\downarrow,2\rangle$ & 0.9957 \\
			$|\uparrow,1\rangle$ & $|\uparrow,2\rangle$ & 0.9977 \\
			$|\uparrow,2\rangle$ & $|\uparrow,1\rangle$ & 0.9977 \\
			$(|\downarrow\rangle+|\uparrow\rangle)\otimes|1\rangle/\sqrt{2}$ & $(|\downarrow,1\rangle+|\uparrow,2\rangle)/\sqrt{2}$ & 0.9988 \\
			\bottomrule[1.5pt]
		\end{tabular}\label{tab1}
	\end{table}	
\begin{figure}[!ht]
	\centering
	\includegraphics[width=1\linewidth]{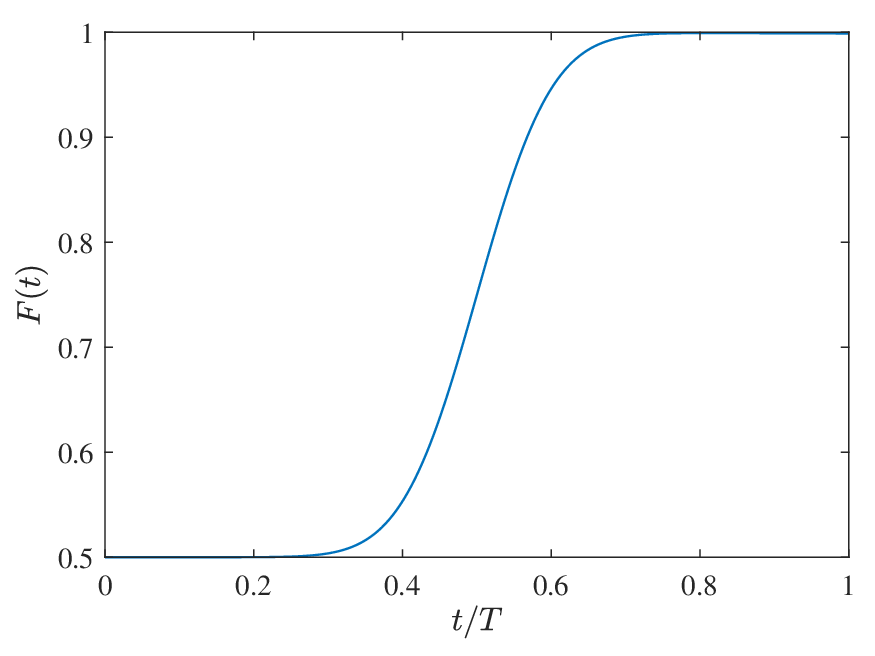}
	\caption{Time evolution of the state $\rho(t)$ under the CNOT gate with the initial state $(|\downarrow\rangle+|\uparrow\rangle)\otimes|1\rangle/\sqrt{2}$. $F(t)$ is the fidelity between the ideal output state and the state $\rho(t)$. The decay rates are the same as Tab.~\ref{tab1}. The duration $T$ of the Gaussian driving pulse is 25~ns. }\label{fig3}
\end{figure}

In Tab.~\ref{tab1}, we present the fidelity of the CNOT gate with typical initial states under the influence of dissipation. The decay rates are calculated under a typical electric holding field $E_{\perp}=100$~V/cm \cite{konstantinov2008ltp}. Because the adiabatic condition is not required during the evolution, the evolving time can be very short. For the typical microwave driving with Rabi frequency $\Omega_R/2\pi=40$~MHz \cite{konstantinov2008ltp, kawakami2021prl}, we use the Gaussian driving pulse with the duration $T=2\pi/\Omega_R=25$~ns and the standard deviation $\sigma=T/8$. The full width at half maxima (FWHM) is $t_{\textrm{FWHM}}=2\sqrt{2\ln 2}\sigma\approx0.3T$.

\begin{figure}[!ht]
		\centering
		\includegraphics[width=1\linewidth]{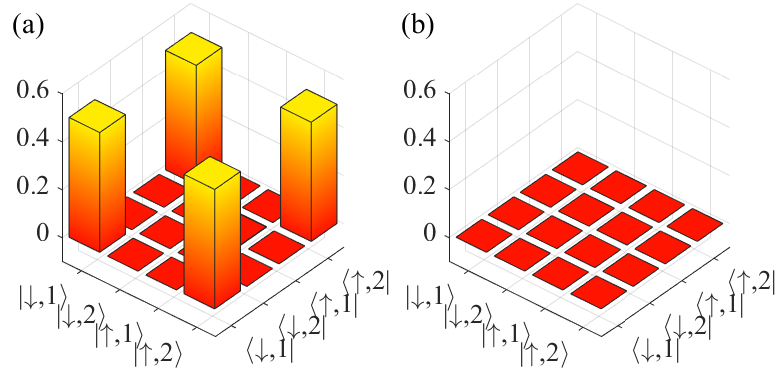}
		\caption{(a) Real and (b) imaginary part of the density matrix of the output state from the initial state $(|\downarrow\rangle+|\uparrow\rangle)\otimes|1\rangle/\sqrt{2}$. The decay rates are the same as Tab.~\ref{tab1}. }\label{fig4}
	\end{figure}

It is noteworthy that the CNOT gate can generate an entangled state from a product state. Thus, we present the time evolution of the initial product state  $(|\downarrow\rangle+|\uparrow\rangle)\otimes|1\rangle/\sqrt{2}$ in Fig.~\ref{fig3} and the density matrix of the final state in Fig.~\ref{fig4}. The result shows that the initial state evolves to the maximal-entangled state $(|\downarrow,1\rangle+|\uparrow,2\rangle)/\sqrt{2}$ with high fidelity. Figure~\ref{fig3} also indicates that the fidelity $F>0.99$ as long as $t>0.67~T$. Even if the pulse duration is a little bit longer or shorter than $T$, the fidelity of the final state is still very high. Thus, the NHQC method is not sensitive to the fluctuation of the pulse duration, which might be commonly observed in experiments.

\begin{figure}[!ht]
		\centering
		\includegraphics[width=1\linewidth]{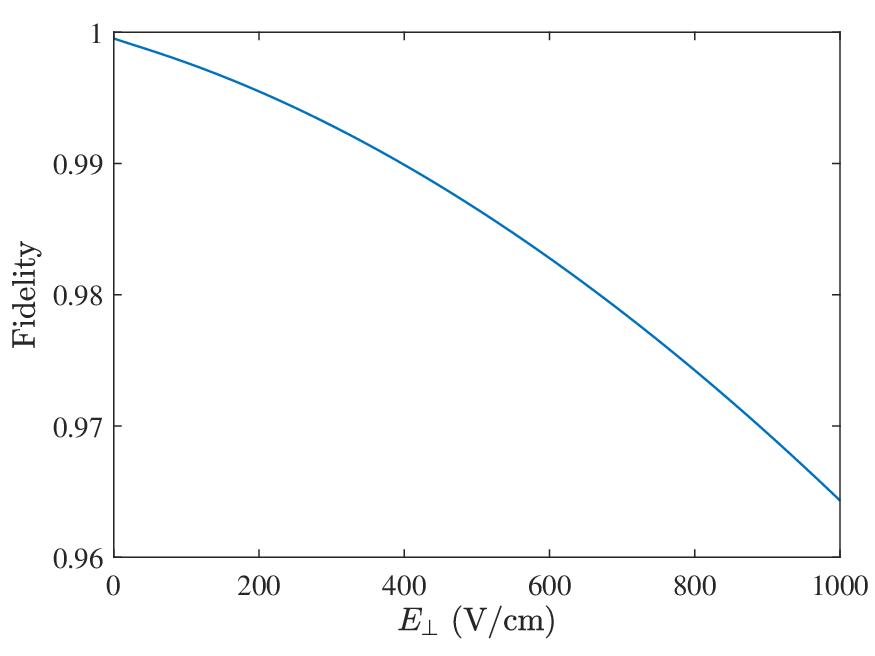}
		\caption{Fidelity of the output state with initial state $(|\downarrow\rangle+|\uparrow\rangle)\otimes|1\rangle/\sqrt{2}$ in different electric fields $E_{\perp}$. }\label{fig5}
	\end{figure}

Generally the electric holding field $E_{\perp}$ is applied to the experimental system in order to confine the motion of electrons and tune the energy spacing between the Rydberg states. $E_{\perp}$ is on the order of $100\sim1000$~V/cm for typical experimental configuration \cite{zou2022njp}.  Because $\kappa_{mn}$ increases with $E_{\perp}$, we investigate the fidelity with initial state $(|\downarrow\rangle+|\uparrow\rangle)\otimes|1\rangle/\sqrt{2}$ under different electric fields, as shown in Fig.~\ref{fig5}. The fidelity $F$ is higher than $0.99$ for $E_{\perp}<400$~V/cm, and higher than $0.96$ for $E_{\perp}<1000$~V/cm.  Therefore, our scheme is robust against dissipation in experiments.

\begin{figure}[!ht]
	\centering
	\includegraphics[width=1\linewidth]{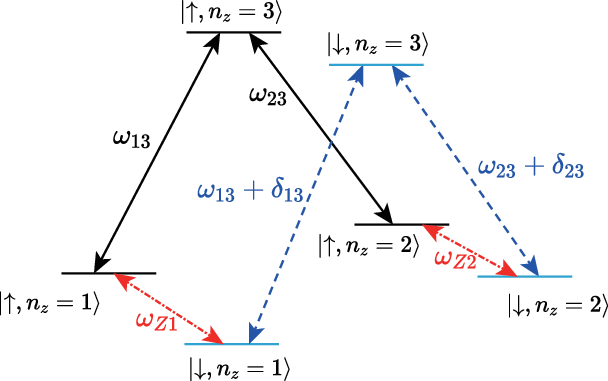}
	\caption{The transition frequencies between different energy levels. In the single-qubit gate proposal, we apply four resonant drivings with frequencies being respectively $\omega_{13}$, $\omega_{23}$, $\omega_{13}+\delta_{13}$, and $\omega_{23}+\delta_{23}$. In the controlled gate proposal which regards the Rydberg states as the control qubit, one resonant driving with frequency being $\omega_{Z2}$ is applied.  }\label{fig6}
\end{figure}

As for the single-qubit gate, we can apply four resonant drivings with frequencies being respectively $\omega_{13}$, $\omega_{23}$, $\omega_{13}+\delta_{13}$, and $\omega_{23}+\delta_{23}$, and simultaneously perform two NHQC gates in the spin up and spin down subspace, as shown in Fig.~\ref{fig6}. In this way, a single-qubit gate on the Rydberg state is performed, i.e.,
\begin{align}
	U_{\textrm{tot}}=|\downarrow\rangle\langle\downarrow|\otimes U +|\uparrow\rangle\langle\uparrow|\otimes U=I\otimes U,
\end{align}
where $U$ is the single-qubit operation in Eq.~(\ref{eqU}). This proposal still works when the driving pulses $\omega_{13}$, $\omega_{23}$ and $\omega_{13}+\delta_{13}$, $\omega_{23}+\delta_{23}$ are not applied simultaneously. In Tab.~\ref{tab2} we present the average output-state fidelity of the single-qubit NOT gate and Hadamard gate. The ``non-simultaneous case'' implies that the driving pulses with frequencies $\omega_{13}+\delta_{13}$ and $\omega_{23}+\delta_{23}$ are applied $T/4$ later than the driving pulses with frequencies $\omega_{13}$ and $\omega_{23}$. The state fidelity is obtained by the following procedures. We begin with the initial state $\rho_{i}^{\textrm{tot}}$, which is the product state of the spin state and the Rydberg state, i.e., $\rho_{i}^{\textrm{tot}}=\rho_{i}^{\textrm{S}}\otimes\rho_{i}^{R}$. Then we derive the final state $\rho_{f}^{\textrm{tot}}$ from the evolution determined by the master equation (\ref{eq7}). Next, we obtain the reduced density matrix $\rho_{f}^{R}$ of the Rydberg state by taking partial trace on $\rho_{f}^{\textrm{tot}}$, i.e., $\rho_{f}^{\textrm{R}}=\textrm{Tr}_{\textrm{spin}}\big(\rho_{f}^{\textrm{tot}}\big)$. Finally, we acquire the state fidelity between $\rho_{f}^{R}$ and the ideal final state. The average fidelity in Tab.~\ref{tab2} is derived by averaging the results of six input states, with the initial spin state being $(|\downarrow\rangle+|\uparrow\rangle)/\sqrt{2}$ and the initial Rydberg states being $|1\rangle$, $|2\rangle$, $(|1\rangle+|2\rangle)/\sqrt{2}$, $(|1\rangle-|2\rangle)/\sqrt{2}$, $(|1\rangle+i|2\rangle)/\sqrt{2}$, and $(|1\rangle-i|2\rangle)/\sqrt{2}$, respectively. The results indicate that for both the NOT gate and Hadamard gate we can achieve near-unity fidelity.

\begin{table}[!h]
	\centering
	\renewcommand\arraystretch{1.5}
	\tabcolsep=0.2cm
	\caption{Average output-state fidelity of the single-qubit gates in the simultaneous case and non-simultaneous case. The decay rates are the same as Tab.~\ref{tab1}. }
	\begin{tabular}{c c c}
		\toprule[1.5pt]
		Single-qubit gate & simultaneously & non-simultaneously \\
		\midrule[1pt]
		NOT gate & 0.9984 & 0.9980 \\
		Hadamard gate & 0.9985 & 0.9981 \\
		\bottomrule[1.5pt]
	\end{tabular}\label{tab2}
\end{table}

\section{Conclusion and remarks}\label{sec5}

In this work, we present a scheme to realize nonadiabatic holonomic gates in an SE system based on the experimental configuration \cite{kawakami2023arxiv}. By applying the state-selective pulses, three Rydberg states with spin up are coupled with driving fields. During the evolution, two orthogonal bases acquire different geometric phases and thus perform a geometric gate. By varying the complex ratio between the Rabi frequencies of the two driving fields, the universal single-qubit nonadiabatic holonomic quantum gate is realized. The controlled-$U$ gate on the Rydberg and spin states is based on the different Zeeman energy splittings in the inhomogeneous magnetic field. With the state-selective driving pulses we perform an arbitrary single-qubit gate $U$ on the Rydberg states with spin up while the Rydberg states with spin down remain unchanged.

It is noteworthy that we can also realize controlled-$U$ gates considering the Rydberg states as the control qubit. As shown in Fig.~\ref{fig6}, the electron-spin-resonance frequencies of $|n_z=1\rangle$ and $|n_z=2\rangle$ are $\omega_{Z1}$ and $\omega_{Z2}$, respectively. Because the magnetic field at the expected positions of $|n_z=1\rangle$ and $|n_z=2\rangle$ are different, the difference between $\omega_{Z1}$ and $\omega_{Z2}$ is $\delta_{12}=\omega_{Z1}-\omega_{Z2}=g\mu_B(B^{(1)}-B^{(2)})$. $\delta_{12}$ is on the order of several hundreds MHz, which is much larger than the decay rate $\sim1$~MHz. Thus, we can resonantly drive the transition between the two spin states and perform a quantum gate through the Rabi oscillation when the Rydberg state is $|n_z=2\rangle$, while keep the spin states unchanged when the Rydberg state is $|n_z=1\rangle$.
	
Because of the fast nonadiabatic evolution, the NHQC proposal is robust against dissipation. Our theoretical scheme is based on the experimental configuration, and the parameters are experimentally achievable. Therefore, this work will supply heuristic insight for fast-manipulation tasks of holonomic quantum computation that involve both the Rydberg and spin states of the SE. 
	
\begin{acknowledgements}
	This work is supported by the National Natural Science Foundation of China under Grant No.~61675028. Qing Ai is supported by the Beijing Natural Science Foundation under Grant No.~1202017 and the National Natural Science Foundation of China under Grant Nos.~11674033, 11505007, and Beijing Normal University under Grant No.~2022129.
\end{acknowledgements}	
	
	\appendix	
	\section{Wavefunction in the external field} \label{AppA}
	
	\begin{figure}[!ht]
		\centering
		\includegraphics[width=1\linewidth]{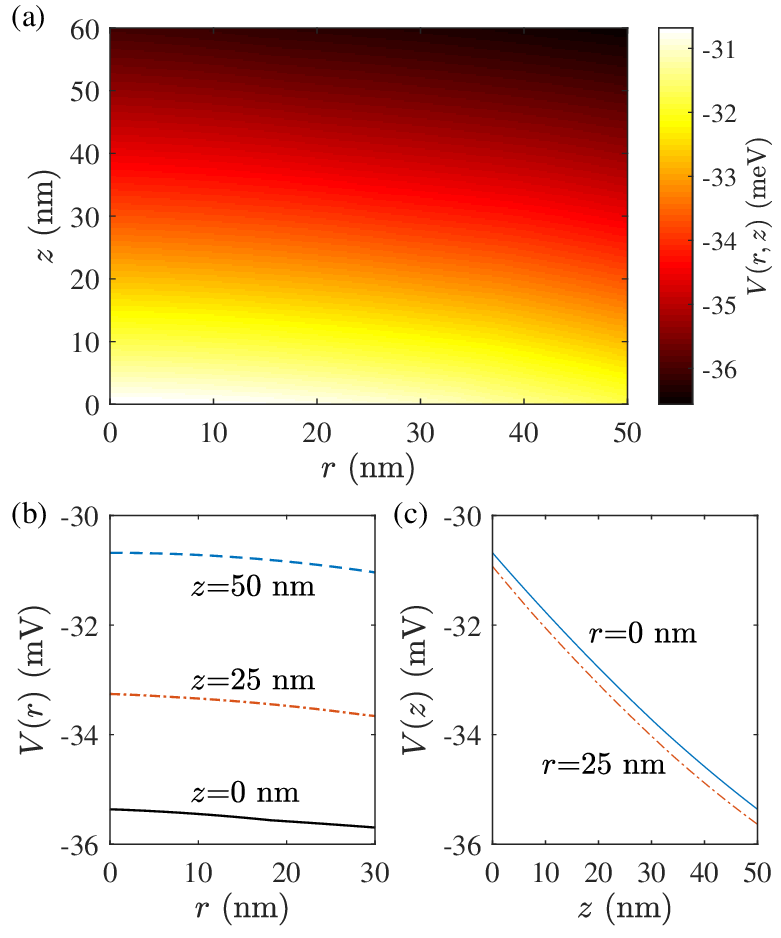}
		\caption{(a) The electric potential $V(r,z)$. (b) The radial electric potential with different vertical coordinates. (c) The vertical electric potential with different radial coordinates. }\label{figA1}
	\end{figure}

	\begin{figure}[!ht]
		\centering
		\includegraphics[width=1\linewidth]{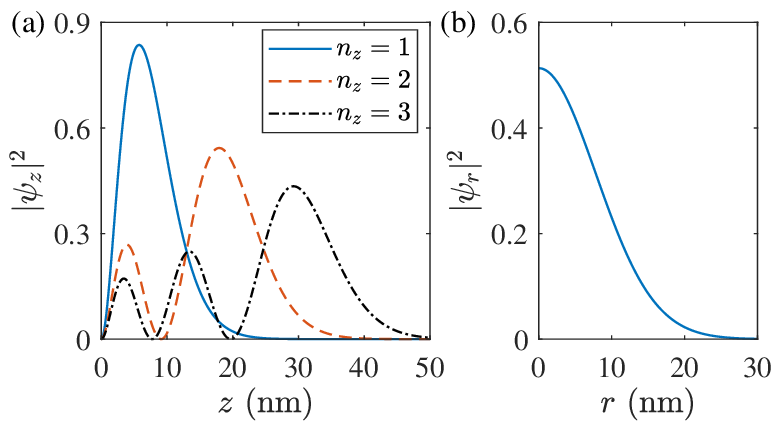}
		\caption{(a) The vertical wavefunction $\psi_z(z)$ with $m=0,~n_r=0$. (b) The radial wavefunction $\psi_r(r)$ with $m=0$, $n_r=0$ and $n_z=1$. }\label{figA2}
	\end{figure}

The electric potential distribution $V(r,z)$ is derived from the finite element analysis, as shown in Fig.~\ref{figA1}. In a large range of $0<z<50~\textrm{nm}$ and $0<r<25~\textrm{nm}$ that the wave function with $n_z=1,2,3$ and $m=0,n_r=0$ are mostly distributed in,  $V(r,z)$'s with different $z$'s or different $r$'s are approximately parallel. Thus, the electric potential can be approximately divided into two parts, i.e., $V(r,z)=V_r(r)+V_z(z)$. The wave function can be written as $\psi(r,z)=\psi_r(r)\psi_z(z)$, where $\psi_r(r)$ and $\psi_z(z)$ are determined by
	\begin{align}
		\left[-\frac{\hbar^2}{2m_e}\frac{\partial^2}{\partial z^2}-\frac{\Lambda e^2}{z}-eV_z(z)\right]\psi_z=E_z\psi_z,
	\end{align}
and
	\begin{align}
		&\Big[-\frac{\hbar^2}{2m_e}\Big( \frac{\partial^2}{\partial r^2}+\frac{1}{r}\frac{\partial}{\partial r}-\frac{m^2}{r^2} \Big) +E_z+m\hbar\omega_c \nonumber \\ &+\frac{1}{8}m\omega_c^2r^2-eV_r(r)\Big] \psi_r=E_r\psi_r,
	\end{align}
respectively. The numerical solutions are shown in Fig.~\ref{figA2}, which indicate that the position of the electrons with $n_z=1,2,3$ and $m=0,n_r=0$ are confined to $0\sim50$~nm along the vertical direction and to $0\sim25$~nm along the radial direction. The expected positions $\langle z_n\rangle$ of the three lowest Rydberg states are respectively 7.63~nm, 17.2~nm, and 25.3~nm.

	
	

\section{Decay mechanism of the surface electron}
\label{AppB}

The two-ripplon scattering in the short-wavelength range dominates the decay mechanism at low temperature. The lifetime of the Rydberg states is on the order of 1~$\mu$s, which is much shorter than the spin relaxation time (50~ms) and the spin dephasing time (100~s) \cite{lyon2006pra,kawakami2023arxiv}. Thus, we only consider the dissipation of the Rydberg states. The decay rate of the first excited state, i.e., $n=2$, is expressed as \cite{monarkha2006ltp, monarkha2007jltp, monarkha2010ltp}
\begin{align}
	\kappa_{12}=\frac{m_e\kappa_0^2}{4\pi\hbar\alpha\rho}\left(\frac{\rho}{4\hbar^2\alpha}\right)^{1/3}\left(\frac{\partial V}{\partial z}\right)_{11}\left(\frac{\partial V}{\partial z}\right)_{22}\Delta_{21}^{2/3},
\end{align}
where $V(z)$ is the electron potential, $\kappa_0$ is the penetration depth of the electron wave function into liquid, $\alpha$ is the surface tension of liquid helium, and $\rho$ is the liquid mass density. The decay rate is determined by the diagonal matrix element of $\partial V/\partial z$ and the energy difference $\Delta_{mn}=E_z^{{(n)}}-E_z^{{(m)}}$ between the initial state and the lower-lying Rydberg states that the initial state leaks to, i.e., the decay rate between Rydberg states $|m\rangle$ and $|n\rangle$ is
\begin{align}
	\kappa_{mn}=\frac{m_e\kappa_0^2}{4\pi\hbar\alpha\rho}\left(\frac{\rho}{4\hbar^2\alpha}\right)^{1/3}\left(\frac{\partial V}{\partial z}\right)_{mm}\left(\frac{\partial V}{\partial z}\right)_{nn}\Delta_{nm}^{2/3}, \label{eqA2}
\end{align}

\begin{figure}[!ht]
	\centering
	\includegraphics[width=1\linewidth]{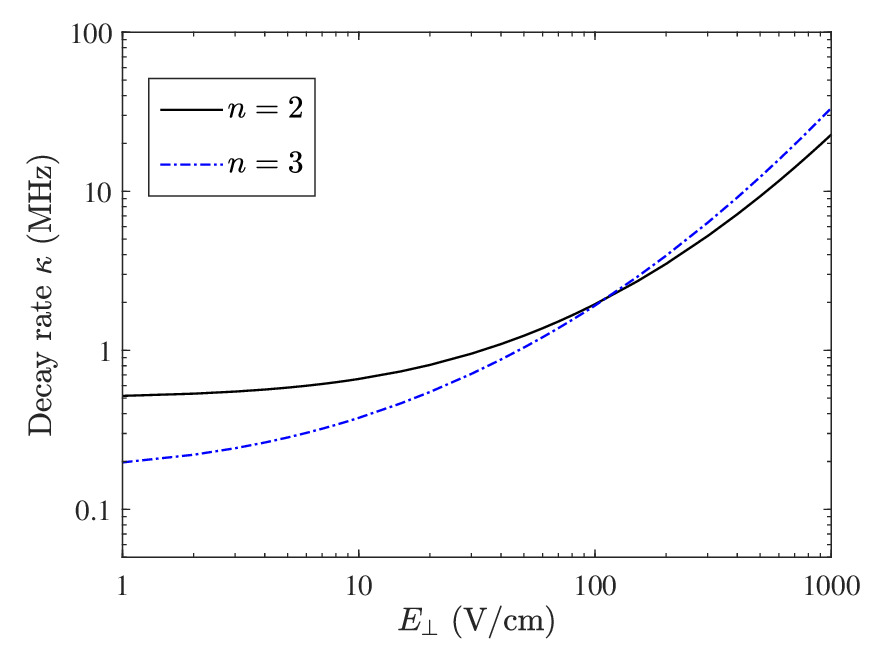}
	\caption{The decay rate of the first and second excited Rydberg state of the SE versus the holding field intensity $E_{\perp}$. }\label{figA3}
\end{figure}

As shown in Fig.~\ref{figA1}(c), the electric holding field is approximately linear along the vertical direction $z$. Thus, we consider the electric holding field $E_{\perp}$ as a uniform field. The corresponding potential can be expressed as $eE_{\perp} z$. For typical experimental configuration, $E_{\perp}$ is on the order of $100\sim1000$~V/cm \cite{zou2022njp}. Thus, $eE_{\perp} z$ is comparable with the image potential $-\Lambda e^2/z$. The eigenvalues and wave functions of the corresponding Hamiltonian are numerically solved. Since the spontaneous two-ripplon emission process decreases the energy of surface electrons, we neglect the leakage to the higher-lying Rydberg states when calculating the decay rate. The total decay rate $\kappa^{(n)}$ of the Rydberg state $|n\rangle$ is the sum of the leakage to all lower-lying Rydberg states, i.e., $\kappa^{(n)}=\sum_{m=1}^{n-1}\kappa_{mn}$. As shown in Fig.~\ref{figA3}, the decay rates of the excited states increase with the electric holding fields. This is because the energy level spacing $\Delta_{mn}$ and the derivative of potential $\partial V/\partial z$ enlarge with increasing $E_{\perp}$ . The electron is closer to the liquid surface in the increasing holding field, thus it is more sensitive to the fluctuation of the surface.

%

\end{document}